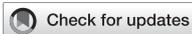

# Connecting space missions through NGSO constellations: feasibility study

Houcine Chougrani[1]*, Oltjon Kodheli[2], Ali Georganaki[1], Jan Thoemel[1], Chiara Vittoria Turtoro[2], Frank Zeppenfeldt[3], Petros Pissias[3], Mahulena Hofmann[1] and Symeon Chatzinotas[1]

[1]Interdisciplinary Centre for Security, Reliability and Trust (SnT), University of Luxembourg, Luxembourg, Luxembourg, [2]Société Européenne des Satellites, SES S.A., Betzdorf, Luxembourg, [3]European Space Agency (ESA), Paris, France

A satellite internet provider (e.g., Starlink, OneWeb, O3b mPOWR), despite possessing the capability to provide internet services to on-ground users in a global scale, can dramatically change the way space missions are designed and operated in the foreseeable future. Assuming a scenario where space mission satellites can access the internet via a space internet system, the satellite can be connected to the network permanently (24 × 7) and act as mere terminal independently from its location. The ability to communicate with the satellite on-demand has the potential to improve aspects such as real-time tasking, outage minimization, operation cost, and dependency on the ground. This paper performs a feasibility study on the concept of connecting space missions to the network through commercial mega-constellations. This study includes a review of existing and near-future space internet systems, identification of candidate space missions for the aforementioned concept, a necessary adaptation of existing Commercial off-the-shelf (COTS) terminals to be plugged into space mission satellites, assessment of communication performance, and investigation of the legal aspects of the radio frequency (RF) spectrum usage. The paper evidences that the concept is practically possible to implement in the near future. Among the studied space internet systems (i.e., Starlink, OneWeb, O3b mPOWER), O3b mPOWER stands out as the most suitable system allowing permanent coverage of low earth orbit (LEO) space missions with data rates that can reach up to 21 Mbps per satellite. Although the concept is very promising and can be implemented in the near future, our investigations show that some regulatory aspects regarding the RF usage should be solved for future exploitation of connecting space missions through NGSO (Non-Geostationary Satellite Orbit) constellations.

# 1 Introduction

Nowdays the number of satellites launched into low earth orbit (LEO), usually for data-gathering missions such as Earth Observation (EO) and remote sensing, is constantly increasing (Euroconsult, 2023). Their satellite operators heavily depend on a ground station, or a network of ground stations, placed at strategic locations to obtain data from these space mission satellites (or send data/command to the satellite). This results in a limited link availability of about once link occasion per orbit (near polar sun-synchronous orbit with polar ground-station or near-equatorial orbit with equatorial ground station) to few per day (near polar orbit with low and mid-latitude ground-station) (Vrancken et al., 2014) this can cost a connection discontinuity up to 90 min until revisit is established.

Increasing the number of ground stations for the purpose of expanding the communication opportunities with the satellite would drastically escalate the cost of a particular LEO mission. With the advent of new type of satellite internet providers (e.g., SpaceX, OneWeb, Amazon) (Al-Hraishawi et al., 2023), communications with space missions satellites can potentially be revolutionized. Thus the paradigm of relying on ground stations at strategic Geo-locations to downlink data from LEO satellites a few times per day can be challenged. Assuming a scenario where LEO satellites (e.g., space missions) can access the internet via a Satellite Internet Provider (SIP), satellites can potentially be connected to the network permanently (24/7). Data can be available on demand, regardless of the satellite's positions. This has the potential to change dramatically the way satellites will be operated. This includes the possibility of the satellite itself initiating an asynchronous (spontaneous) link to the ground (e.g., in case of contingency), and real-time tasking (e.g., collision avoidance, TT&C, data offloading, etc.).

Currently, only a limited number of space-to-space connectivity solutions exist. In the geostationary (GEO) orbit category, the Inter-Satellite Data Relay System (IDRS) is a solution launched by Inmarsat and Addvalue, which is capable of providing on-demand 24/7 connectivity to LEO satellites in all orbital inclinations and at altitudes of up to 1,000 km (Idrsspace, 2023). The data rates supported by their services are 200–300 kbps. In the same category, we have also the National Aeronautics and Space Administration (NASA) Tracking and Data Relay Satellite (TDRS) (NASA, 2023) and European Data Relay Satellite (EDRS) (ESA, 2023). Due to GEO satellites' manufacturing and lunch costs, utilizing the Data Relay Satellite (DRS) systems in the traditional context is limited to niche and private usage with countries demanding their own space DRS to connect their space endeavours. Some of the examples are the Luch of Russia, the Tianlian of China (or CTDRS), Japanese DRS (JDRS), and India's plan for their own DRS (IDRSS). In the LEO category, Kepler Inc.'s satellites are a system that aims to provide real-time communications to other satellites, space stations, launch vehicles, habitats, and any other space-borne assets in the near future (Kepler, 2023). With their new terminals specifically designed for space, the data rate supported by the Kepler system is between 10–40 kbps.

While the range of data rates provided by the above-mentioned systems might be enough for certain applications (e.g., telecommand and control operations), it is insufficient for others where higher throughput is required. Most importantly, the proposed concept in this paper does not require any specific link to connect space missions but the idea herein is to leverage the already existing link for ground users to connect the space missions (see Figure 1). Meaning that a space mission is not technologically required to establish an expensive DRS in GEO designed to serve one particular nation, but to utilize an existing mega-constellation orbiting at a higher altitude. From a SIP perspective, this requires few modifications[1] to their existing system, and space missions, in this context, are seen as mere terminals. This concept was proposed for the first time in our previous work in (Al-Hraishawi et al., 2021), where preliminary analyses were done. And it gained interest in the field as in (Capez et al., 2023; Gabriel Maiolini Capez and Caceres, 2023), where numerous advantages of the concept are demonstrated and the importance of a study on licensing and legal frequency usage is emphasized. To complement the work done previously, this paper provides a deep analysis on the technology needed on space mission satellites, communication link performance as well as legal aspects related to Radio frequency (RF) usage, including both LEO and MEO SIPs. The scope of this work is to focus on existing and near-future systems aiming to offer high-speed broadband services to on-ground users and perform a feasibility study on using the same system to connect space missions. The main contributions of this paper are summarized as follows:

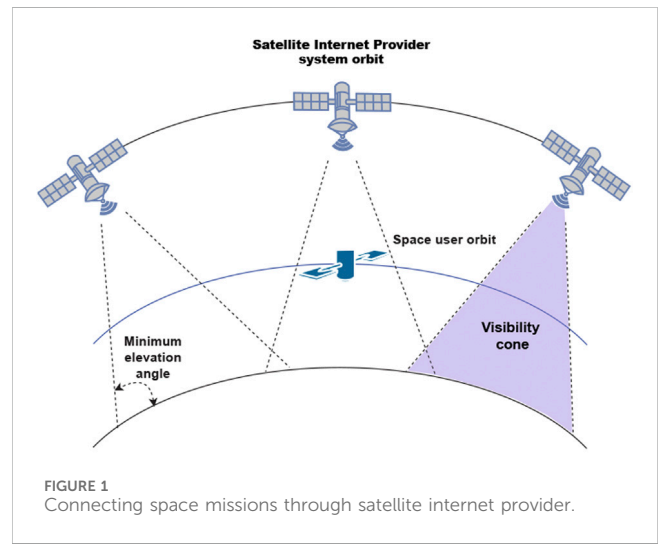

FIGURE 1
Connecting space missions through satellite internet provider.

1. A survey of the main current and near-future SIP with their features, services as well as their orbital characteristics.
2. Identification of the space mission categories that may take advantage of the proposed concept.
3. Investigation of the necessary technologies on-board the space mission-satellite, including the review of existing technologies and identification of potential ones to be integrated into a LEO user-satellite.
4. Assessment of communication performance, including coverage analysis to quantify the effective visibility time to establish a communication link, link budget evaluation, and performance analysis for a hypothetical system composed of the selected space mission and the most promised SIP.
5. Investigation of legal aspects of the RF spectrum usage. This includes the examination of the current spectrum usage policy vis-à-vis the proposed concept, necessary conditions for compliance with the regulatory framework, and recommendations to make the concept viable.

---

1 Modifications on the communication protocol might be needed to deal with the high relative motion of space mission satellites.

TABLE 1 Summary missions benefiting from satellite internet provision and valuation per benefit.

| Mission category\Potential advantage | High data rate Vrancken et al. (2014); Al-Hraishawi et al. (2023); Euroconsult. (2023) | link availability Vrancken et al. (2014); Al-Hraishawi et al. (2023); Euroconsult. (2023) | low latency Vrancken et al. (2014); Al-Hraishawi et al. (2023); Euroconsult. (2023) | Sum |
|---|---|---|---|---|
| Human space flight missions | 3 | 3 | 3 | 9 |
| Experimental/IOD missions | 1 | 1 | 1 | 3 |
| Earth observation missions | 3 | 3 | 1 | 7 |
| Other scientific missions | 1 | 1 | 1 | 3 |

TABLE 2 Size, mass, power characteristics of reference missions.

|  | NASA aqua | ESA BIOMASS |
|---|---|---|
| orbit | SSO 1:30 LTAN, 700 km | SSO 6 a.m./pm, 660 km |
| size [m3] | 40 | 20 |
| mass [kg] | 3,000 | 1,200 |
| power [W] | 5,000 | 1,500 |

The content of the paper is organized as follows. Section 2 describes the system elements including space missions and SIPs with their features; Section 3, presents the results of the feasibility study of the proposed concept; Section 4 is dedicated to spectrum usage policy investigation; finally, Section 5 provides the conclusions of this work.

## 2 System elements

### 2.1 Space missions

To understand the added value of SIPs to space missions, we performed a detailed review and categorization of past, ongoing, and future space missions. After eliminating interplanetary missions and Pico/Nanosatellite due to their apparent incompatibility with SIP enforced by their location in space, system performance, and size, we constrained our research to the rest of space missions that can reasonably receive the nadir beams of SIP satellites residing in LEO and MEO, respectively. Depending on their need and requirements (e.g., throughput, link availability, and latency.) we classify them into four categories.

- Human space flight currently consists of the International Space Station (ISS) (Thirkettle et al., 2002) and the Chinese space station Tianhe (Mallapaty et al., 2021), However in the future there is the expectancy of private space stations to be lunched as well. It is worth mentioning the newly born category of commercial space travel which are now in need of massive data collection and transfer to guarantee the safety requirements of future commercial space travels.

The needs of such missions are (Euroconsult, 2023) high data rate due to the production of large amounts of telemetry and science data, (Vrancken et al., 2014), low latency connectivity for voice and video applications, and (Al-Hraishawi et al., 2023) long-duration or permanent connectivity for safety monitoring.

- Experimental missions, such as most CubeSat missions and others that serve in-orbit (IOD) demonstration purposes. These kind of missions are very diverse and typically do not have particular communication needs. They may benefit however from low-cost connectivity.

- EO missions acquire data for the characterization of the Earth's surface and atmosphere and hence produce a large amount of data. These missions are primarily in LEO due to the benefits of the proximity to the observation target with few exceptions in higher orbits. Some of those missions address exploratory fundamental science purposes (Berger et al., 2003). Their data is processed over years after the data acquisition. Other EO missions are of operational nature serving the monitoring of the atmosphere's chemical composition (Fussen et al., 2019; Thoemel et al., 2019). Those need to transfer their data to users with a short delay.

- Scientific missions are typically conducted by space agencies for the scientific community. Examples include missions for astronomical observations. They typically acquire large amounts of data and hence need high-throughput links, but such missions have typically no stringent data availability and latency needs. Thus, they can downlink data upon opportunity such as an available ground station link.

Based on the analysis of space mission communication needs, we estimate that the main potential advantages of connecting space missions to SIP are large data amount transfer (from space mission to satellite internet provider's satellite) and long duration or even permanent connectivity features. However, some missions would benefit from low-latency and low-cost connectivity. As prices cannot be predicted at the time of writing, the latter was omitted in the remainder.

For the first three potential advantages, we estimated the degree of advantage in order to prioritize our research. A simple scoring scheme from one (low importance) to three (high importance) was used. For instance, CubeSat missions normally featuring modest scientific instruments need much less data rate than EO missions such as the EO mission ALTIUS featuring spectral imager. Thus, for CubeSats, often 9,600 baud UHF systems (Mason et al., 2015) suffice while EO mission require minimum an X-band system (Vrancken et al., 2008) capable of transmitting hundreds of Mbaud. The actual relative scoring was estimated based on the needs of our reviewed missions. The scoring of the three potential advantages was equally ranked and then summed up. This is shown in Table 1.

TABLE 3 Space internet provider's constellation parameters.

| SIPs | No. of sat | Altitude (km) | Planes | Sat. per plane | Incl. (°) | Min. Elev. (°) | Lat. (ms) |
|---|---|---|---|---|---|---|---|
| **Starlink** | 4,408 | 540 | 72 | 22 | 53.2 | 25 | ~50 |
| | | 550 | 72 | 22 | 53 | | |
| | | 560 | 6 | 58 | 97.6 | | |
| | | 560 | 4 | 43 | 97.6 | | |
| | | 570 | 36 | 20 | 70 | | |
| **OneWeb** | 6,372 | 1,200 | 36 | 49 | 87.9 | 25 | ~100 |
| | | | 32 | 72 | 55 | | |
| | | | 32 | 72 | 40 | | |
| **Telesat** | 1,671 | 1,015 | 27 | 13 | 98.98 | 10 | - |
| | | 1,325 | 40 | 33 | 50.88 | | |
| **Kuiper** | 7,774 | 590 | 56 | 28 | 33 | 20 | - |
| | | 610 | 72 | 36 | 42 | | |
| | | 630 | 68 | 34 | 51.9 | | |
| | | 640 | 652 | 1 | 72 | | |
| | | 650 | 325 | 2 | 80 | | |
| **O3b mPower** | 60 | 8,062 | 1 | 44 | 0 | 5 | ~150 |
| | | | 2 | 8 | 70 | | |

Our scoring shows that human space flight and Earth observation missions are the likely missions that benefit most from SIP connectivity.

Accounting for the abundance of EO missions, their overwhelming need for broadband data rate, and the rarity of human space flight missions, we concluded that the first users of SIP connectivity are likely EO missions. Hence, we chose EO missions for the study and selected NASA's past mission Aqua and ESA's future mission BIOMASS as a reference. Their main features are given in Table 2.

## 2.2 Satellite internet providers

In this section, we describe the main technical peculiarities of current and near-future SIPs that are relevant to the proposed concept. In the LEO category, we investigate four of the largest mega-constellations currently providing, or envisioned to provide broadband services, namely, Telesat, OneWeb, Starlink, and Kuiper. In the MEO category, we consider O3b mPOWER. We did not consider the GEO category, even though it is also a candidate for the proposed concept[2]. Table 3 summarizes the space internet provider's constellation parameters.

---

[2] The GEO systems might provide good performance but their main weaknesses are the absence of coverage in the polar regions and higher latencies compared to NGSO solutions.

### 2.2.1 Starlink

On November 2020 SpaceX announced its testing phase of offering internet services, termed as "better than nothing (beta)" (SpaceNews, 2023). According to the latest reports, the testing phase ended in October 2021 (Teslarati, 2023), and the broadband services offered by Starlink are available nationwide, although still limited by the peak number of users (Boris, 2023). As stated in their latest Federal Communications Commission (FCC) filings approved on April 2021 (FCC, 2023d), Starlink mega-constellation will be mainly composed of 4,408 satellites placed in five orbital shells. The first shell has already been deployed and constitutes the initial phase, orbiting at 550 km altitude. The second phase involves the deployment of the 4 other orbital shells orbiting at altitude range of 540 km–570 km, all operating with a minimum elevation angle of 25°. As of 2023, Starlink provides focused availability to the USA, Canada, Europe, UK, Japan, Australia, and some parts of Latin America and Africa household users, with a mean download speed of 100 Mbps and upload speed of 20 Mbps for 120 $/mo plus equipment price, and mean latency of around 50 ms (PCMAG, 2023).

### 2.2.2 OneWeb

According to the FCC filings (FCC, 2023c; FCC, 2023b; FCC, 2023a), OneWeb's system mainly consists of two phases. Phase 1 was initiated on April 2016 (FCC, 2023c), when OneWeb requested the authorization to operate 720 non-geostationary (NGSO) satellites at a 1,200 km altitude. On May 2020, a modification for the Phase 1 was requested, reducing the number of satellites from 720 to 716, and modifying the orbital planes of some of the satellites in the constellation (FCC, 2023b). In order to reach global coverage, on January 2021, OneWeb requested to FCC the authorization to operate 6,372 satellites for its Phase 2 (FCC, 2023a). The second phase intends to drastically increase the number of satellites in the constellation by increasing the number of the already existing planes in Phase 1, plus an additional third set of orbital planes inclined with 40°. Similar to Starlink, minimum elevation for the service link of OneWeb is 25°. As of 2023, OneWeb who merged with Eutelsat is primarily focused on the European and Asian markets with solutions for maritime, aviation, governments and land mobility with no publically advertised pricing. In principle OneWeb satellites can provide service to all Earth locations, specially the poles with latency less than 100 ms (OneWeb, 2023).

### 2.2.3 Telesat

As described in their FCC filings (FCC, 2016b; FCC, 2020), Telesat's system will consist of 298 satellites in the initial phase, and reach 1,671 satellites in the final phase. Ka-band frequencies will be utilized for both user and GW links, and the final constellation will operate with a minimum elevation angle of 10°. Two types of orbits will be used, high-inclined or polar orbits at 1,015 km, and low-inclined orbits at 1,325 km, with minimum elevation angle of 10°. As of 2023, Telesat Lightspeed constellation provides global coverage with 198 satellites for data and Telecom, mobility, government, and video (Telesat, 2023).

### 2.2.4 Kuiper

Kuiper is the newest mega-constellation system announced by Amazon in 2019 to offer broadband services. In their initial FCC filings which has already been approved, it has been requested to build a system containing 3,236 satellites (FCC, 2019) with 20° as minimum elevation angle for both service and GW links. The satellites will be distributed in three orbital shells at altitudes of 590, 610, 630 km.

Recently, in November 2021, a new request has been submitted to FCC to launch another 4,538 satellites, expanding the number of satellites to 7,774 (FCC, 2021). And also doubling the initial orbital planes, thus doubling the number of satellites with the same orbital configurations. Also, two orbital shells are introduced, separated by 20 km in altitude one with 652 orbital planes at 72° inclination containing 1 satellite each at 640 km altitude, and another one with 325 orbital planes at 80° inclination containing 2 satellites each at 650 km altitude As of the license from the FCC, half of the Kuiper constellation must be launched by 2026, and the full constellation by 2029.

### 2.2.5 O3b mPOWER

O3b mPOWER is the future satellite constellation in Medium Earth orbit (MEO) owned and operated by SES (SES, 2023b). The O3b mPOWER system targets markets as cruise ships, off-shore energy, cloud-scale IP networks, Intelligence-Surveillance-Reconnaissance, enterprise-level organizations, mobile backhaul, commercial shipping, government network solutions, commercial aero, and generic cloud services (SES, 2023d). According to their FCC filings (FCC, 2016a), the final constellation will consist of two orbital planes with a 70° inclination containing 8 satellites each and 44 satellites in the equatorial plane with a minimum elevation angle of 5°. According to (SES, 2023a), the O3b mPOWER constellation will consist initially of 11 powerful satellites, each equipped with more than 5,000 digitally formed beams, and an extensive next-generation O3b mPOWER ground infrastructure. The first two O3b mPOWER satellites were successfully launched in December 2022. It is worth noting that SES has currently an O3b constellation composed of 20 satellites, orbiting in an equatorial plane. This constellation is not O3b mPOWER but can be seen as a first generation of it. O3b mPOWER can provide full Earth coverage with latency of 150 ms (SES, 2023c).

## 2.3 Latency aspects

The guaranteed latency by currently active SIPs in LEO and MEO have mean value of 50 and 150 ms respectively. It is safe to say that a space mission will experience a few millisecond less delay with respect to a ground user depending on its operating altitude. For instance, a LEO space mission at 300 km (resp. 660 km) altitude will have $\approx$ 4.3 ms (resp. $\approx$ 8.8 ms) less latency compared to a ground user. For example, Biomass mission at 660 km altitude connecting to OneWeb (resp. O3b mPOWER) will have $\approx$ 91 ms (resp. $\approx$ 140 ms) latency, assuming 100 ms (OneWeb, 2023) and 150 ms (SES, 2023c) latency for OneWeb and o3b mPOWER constellations, respectively.

Typically, EO missions generate large amounts of data, which is downlinked to polar ground stations within 3 h as a requirement in Near Real-Time products (Copernicus. Sentinels, 2023) (or 48 H in Slow-Time Critical or 1 month - 1 year in Non-Time Critical products). Hence low latency is not crucial for the current missoins; however, the data rate offered by the concept enables much faster data delivery with more availability compared to every 90 min polar revisit (link availability latency) which also alleviates the memory requirement onboard the payload. This can improve services such as weather modelling and forecast, and can potentially introduce Real-Time products for future EO missions. On the other hand, human space flight missions require permanent communication link to monitor the general health of the spacecraft and the crew, and a low latency link serves audio communication between astronauts and the ground. We conclude that connection through SIP, with low latency and permanent communication, not only fulfills space mission requirements but also could potentially address new cases and services in the future.

# 3 System feasibility analysis

## 3.1 Space user terminal

To aid the research for the use of SIP, an investigation is carried out to determine the necessary technologies on-board the user-satellite (i.e., space mission). A system engineering approach is adopted. A review of existing technologies is carried out helping to understand their potential and compatibility with the proposed concept. Then a buy-make-modify decision is made. In this context, NASA conducted a study on the use of broadband satellite provider (Kul et al., 2020) concluding that no terminals are immediately available for the use on their missions. Currently, Thinkom advertises a development regarding their terminal use on satellites on their website (Thinkom, 2023a). GetSat states informally that no activities are planned in this regard due to the limited market size.

TABLE 4 Available COTS terminals and their key features.

| Product [freq. Band] | Max. EIRP [dbW] | G/T [dB/K] | Elev | link acquisition [s] | Steering method | Mass [kg] | Size [cm] |
|---|---|---|---|---|---|---|---|
| Nighingale I [ka] CESIUM. (2023) | 30 | n/a | 30-90 | n/a | ESA | 1.2 | 12 × 8.3 × 1 |
| Nanosat [Ka] GETSAT. (2023c) | 39.7 | 2.2 | 0-90 | < 30 | MSA | 2.3 | 19.5 × 16 |
| Microsat LW [Ku] GETSAT. (2023a) | 42.7 | 4.2 | 0-90 | < 30 | MSA | 4 | 29 × 21 |
| Microsat LW [Ka] GETSAT. (2023a) | 46.3 | 8.2 | 0-90 | < 30 | MSA | 3.2 | 29 × 21 |
| Millisat W LW [Ku] GETSAT. (2023b) | 46.1 | 7.2 | -8-90 | < 30 | MSA | 17.7 | 57.5 × 27 |
| Millisat W LW [Ka] GETSAT. (2023b) | 49 | 11.2 | -8-90 | < 30 | MSA | 17.3 | 57.5 × 27 |
| Sling Blade LM milli [Ku] GETSAT. (2023d) | 46.5 | 11.5 | 10-90 | < 30 | ESA | 48 | 66 × 9 × 90.9 |
| Micro Sling Blade LM milli [Ka] GETSAT. (2023a) | 49 | 10.5 | 10-90 | < 30 | ESA | 18.7 | 60 × 7 × 56 |
| u8 GEO [Ku] Groundcontrol. (2023) | 45.5 | 11.5 | 15-90 | n/a | ESA | 32 | 90 × 90 × 12 |
| ThinPack Ku100 [ku] ThinKom. (2023b) | 40 | 11.5 | n/a | n/a | Hybrid | 2.7 | 40 × 25 × 6 |
| ThinPack Ka100 [ka] ThinKom. (2023a) | 46 | 13 | n/a | n/a | Hybrid | 4.2 | 41 × 28 × 5 |

Terrestrial solutions exist to connect to broadband provider satellites in geostationary, mid-altitude and low-altitude orbits. They:

- Are very commonly available commercial products
- Have a very high maturity as the products are delivered as turnkey solutions to private and governmental customers
- Are designed for on-the-move applications (e.g., boats, planes and helicopters) and under harsh environments
- Provide high reliability uses such as for combat.

Commercial off-the-shelf (COTS) terminals are fully integrated and can often be used without major set up. The user has the choice to use the integrated modem and its Transmission Control Protocol/Internet Protocol (TCP/IP) interface or alternatively an Open Antenna to Modem Interface Protocol (OpenAMIP) (Idirect, 2023) interface to control the antenna. The latter has emerged as an industry standard for such applications.

In particular, we have considered for this study: (Euroconsult, 2023): GetSat's mechanically steerable antennas (MSA) of the Nanosat, Millisat and Microsat series, (Vrancken et al., 2014), GetSat's electronically steerable antennas[8] of the Sling Blade series, (Al-Hraishawi et al., 2023), Kymeta's u8 terminal and (Idrsspace, 2023) Thinkom's Thinpack series with hybrid steerable antennas (HSA). All the relevant terminals are summarized in Table 4 with their key features.

The unprecedented use of a broadband terminal on-board a satellite impacts the satellite as well as the terminal. For either, we carried-out a delta-design analysis, i.e., we assessed what aspects are different in such use-case over the previous uses. We then found that the suitability of broadband terminals for LEO space mission satellites needs considerations in three main criteria: (Euroconsult, 2023): communication function, that is, the interface to the host satellite and to the mega-constellation satellite internet providers, (Vrancken et al., 2014), space and launch environment, and (Al-Hraishawi et al., 2023) satellite system engineering resources availability.

### 3.1.1 Communication function

Terrestrial terminals are typically highly integrated and feature all necessary functions to establish a link to a broadband satellite providers. As such they are optimized to establish and maintain a strong link and minimize latency. To this end, they employ features such as:

- Fast steering of the antenna (either MSA, ESA or HSA). This is made possible due to location awareness, thanks to an integrated global navigation satellite systems (GNSS) system, the stored ephemerides of provider satellites in combination with an inertial measurement unit (IMU), and the received signal strength indicator (RSSI) tracking method.
- Use of standard interfaces such as point-to-point TCP/IP through ethernet from integrated modem, and OpenAMIP.
- Power amplifier/block-up-converter
- Commonly used frequency bands such as Ku and Ka-bands
- Cold and hot start connection capability as well as hand-over between satellites or providers.

The communication function is hence mature and requires no major engineering for its use on LEO satellites.

---

8 Hereafter "ESA".

### 3.1.2 Space and launch environment

The use of terrestrial terminals in space poses a challenge with respect to their compatibility to the new environment, and in particular for:

- Mechanical loads, in especially launch vibrational loads. The current use-case of the terminals for on-the-move applications implies a high mechanical resilience. A particularity is the steering mechanism of the MSA, which is sensitive to mechanical loads, which, if used, would need to be redesigned; inherently, ESAs are less sensitive due to the absence of mechanisms.
- Material use. Specifications of the investigated terminals show the use of plastics as covers. For space-use, such would be prohibitive and, in most cases, not necessary. Antennas are maintenance-free implying encapsulated lubrication of mechanisms. Yet, a detailed review-of-design of material used needs to be conducted to assess the compatibility to LEO vacuum and atomic oxygen environment. This will likely result in the removal of plastic materials.
- Thermal loads. Terminals dissipate a significant amount of heat through the use of fans - not feasible in space. For use in the vacuum of space, a thermal management and cooling concept is needed ensuring the transport heat away from the terminal and and radiate heat into space. We estimate that heat conduction with metal conductors or heat pipes and classical radiators suffice this purpose.
- Electromagnetic interference. The COTS terminals' current use-case scenario is generic and hence indicates a robust electromagnetic interfernce (EMI) design to incoming irradiation. The terminal-radiated EMI, however, is likely significant due to the large emitted RF power levels, which may impact the host satellite. Furthermore, the current use-case (i.e., on-ground use) may be agnostic to radiation lobes at the back surface, which may not be the case for on-satellite use. Hence, a detailed analysis test and possibly a mitigation design are needed.
- Particle radiation. The SIPs user satellite orbit at a low altitude and are thus located at the lower end of the van Allen belt where the trapped proton and electron radiation is low. Hence, the effect of radiation on terminal electronics might be low and not impactful. However, the susceptibility to space radiation requires dedicated analysis (Sinclair and Dyer, 2013).

### 3.1.3 Satellite system engineering resources availability

From the main satellite system engineering characteristics of space missions, we can derive what resources would be available for the broadband communication subsystem. Those are:

- Surface magnitude availability (zenith): 1 m² order of magnitude
- Electrical power magnitude availability: hundreds of Watts (orbit average)
- zenith-surface mounted terminal should have permanent zenith visibility.

The last item deserve attention: whereas SIP-satellites are in view of the user-satellite's zenith surface for extended duration, the SIP beam is typically not - due to its relatively small beamwidth, which can be seen in Figure 1. Hence, the "effective" visibility where the communication can be established will depend on the SIP's beams and constellation characteristics (e.g., number of satellites, altitude). Section 3.2 will be dedicated to evaluate the effective visibility time.

The use of SIP requires frequent switches from one SIP satellite to another. This advocates the use of ESA, which can switch with shorter delay from one SIP-satellite to another or employ even several beams simultaneously.

### 3.1.4 Suggested delta design

The investigated terminals listed in Table 4 have a high functional maturity, plus well-defined and proven interfaces. From this point of view, they appear ready for integration into a LEO space mission satellite. Functional verification would suffice. Their consumption of system engineering resources (size, surface, mass, power) is well within what relatively small satellites can provide. However, commercial products are not designed for the space environment. From our analysis, the main design challenge appears to be the thermal management. Commercial terminals rely on advective and fan-enhanced cooling, which is unavailable and infeasible in space. A space-grade thermal management system consisting of heat transport and heat rejection components needs to be added. Materials used may not be compatible with the LEO environment. A detailed design review is needed, and all non-space grade materials need to be removed or replaced. The harsh launch vibrational environment may be detrimental to these commercial products. A review-of-design followed by a shaker-test verification is needed. Possibly, the structure and mechanical interfaces need reinforcements.

From the above analysis, we can conclude that COTS terminals have a great potential to be used on LEO broadband user missions. Their overall maturity allows their integration into satellites after the delta design work on the thermal management and other detailed review-of-design verification has been carried out. Also, a standard system engineering design activity has to be carried out for the intended mission. Among the two major categories of terminals (i.e., MSA and ESA), MSA should be discarded.

From the considered terminals, the GetSat's Sling Blade and Thinkom's ThinPack Ka/Ku100 stand out as good candidates for space mission satellites.

## 3.2 Coverage analysis

In this sub-section, we perform the coverage analysis for various SIPs using the System Tool Kit (STK) modeling environment (AGI STK version 11.5.0). Two main metrics have been evaluated, the coverage time from the constellation, and the coverage time from each individual satellite. We select three SIPs for the evaluation, Starlink, OneWeb, and O3b mPOWER. The coverage of the other SIPs (e.g., Kuiper, Telesat) can be concluded by analogy, since their orbits are similar to either Starlink or OneWeb. In this analysis, we assume the same space segment characteristics of the SIP (i.e., steering capabilities of the satellite antenna) which are derived based on the minimum elevation angle that SIP systems can reach on the ground.

Regarding the space missions, as highlighted in Section 2.1, we have selected Aqua and Biomass missions for the coverage analysis

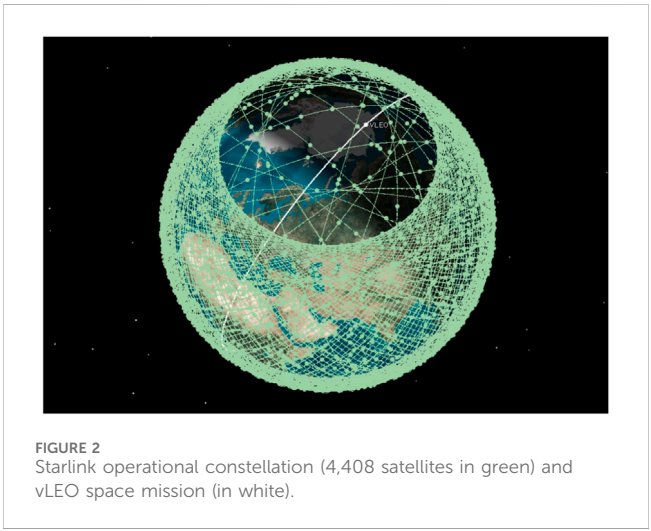

**FIGURE 2**
Starlink operational constellation (4,408 satellites in green) and vLEO space mission (in white).

(shown in Figures 3, 4). It is worth highlighting here that for Starlink the range of satellite altitudes is at least 100 km lower than the altitude of the selected space missions (i.e., Aqua with perigee altitude of 702 km and Biomass at 660 km). Therefore, Starlink constellation might be suitable only for future missions in the very-LEO (vLEO) category where the range of altitudes can be from 200–500 km (Llop et al., 2014). Therefore, the analysis for Starlink is done by taking into account a hypothetical futuristic vLEO mission at 300 km altitude and sun-synchronous orbit (shown in white color in Figure 2.

### 3.2.1 Constellations modelling

Starlink and OneWeb are composed of a large number of satellites, 4,408 and 6,372, respectively. Therefore, to ensure reliable and reproducible results we have considered full operational constellation of Starlink consisting of 4,408 satellites shown in Figure 2, and 630 orbiting satellites of OneWeb Figure 3 which concluded Phase 1 launches. These full constellations were imported from STK Database on October 2023. However, The O3b mPOWER constellation used in this work is custom made using FCC information consisting of 60 satellites shown in Figure 4. It is worth noting that the potential sources of error in the used methodology, could be due to the orbit propagation model given by STK, and there might be a slight variation compared to reality.

### 3.2.2 Access duration

We analyzed the access and outage duration from each constellation to the relevant space mission for 1 day. The results are given in Figure 5.

First, for Starlink, as can be noticed in Figure 5A, many interruptions (i.e., outages) occur and the access durations are short. For OneWeb, the situation is better with fewer interruptions and longer access durations (see Figure 5B), while for O3b mPOWER the coverage is permanent and no interruption is observed as shown in Figure 5C. These results are expected since the coverage duration is proportional to the difference in altitude between space mission satellites and SIP ones. This can be seen as a cone (see Figure 1) where its apex is the SIP's altitude and its base is the coverage for the space mission. Hence, assuming a constant angle, the greater the height the larger the base (i.e., coverage).

Table 5 quantifies these results in terms of numbers. The table provides the total duration of access from space missions to the constellation (a set of satellites serving, one after another) together with the minimum and maximum coverage duration without interruptions.

For Starlink, the total access duration given to a vLEO at 300 km is slightly below 20.5 h (i.e. 85.2% per day) with 4.62 s and 35 min as minimum and maximum coverage duration without interruption, respectively. OneWeb's total access duration reaching almost full day coverage (i.e. 99.36%) for Biomass and (i.e. 96.06%) for Aqua, with a minimum and maximum coverage duration without interruption of 22.40 s and 1.5 h, respectively for Biomass. The higher the altitude of the platform (i.e., closer to the SIP) the shorter access durations it will receive. This is show cased by Aqua which is only 40 km higher in altitude with respect to Biomass (inclined equally) and receives less connectivity. O3b mPOWER is the most suitable system with a permanent connection to space missions.

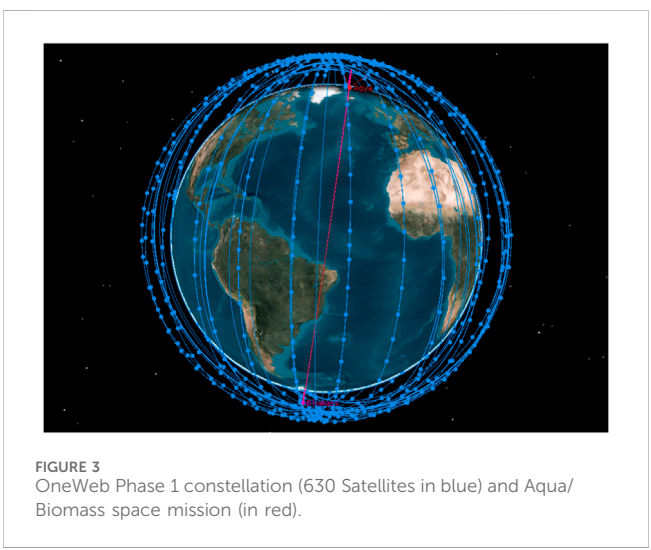

**FIGURE 3**
OneWeb Phase 1 constellation (630 Satellites in blue) and Aqua/Biomass space mission (in red).

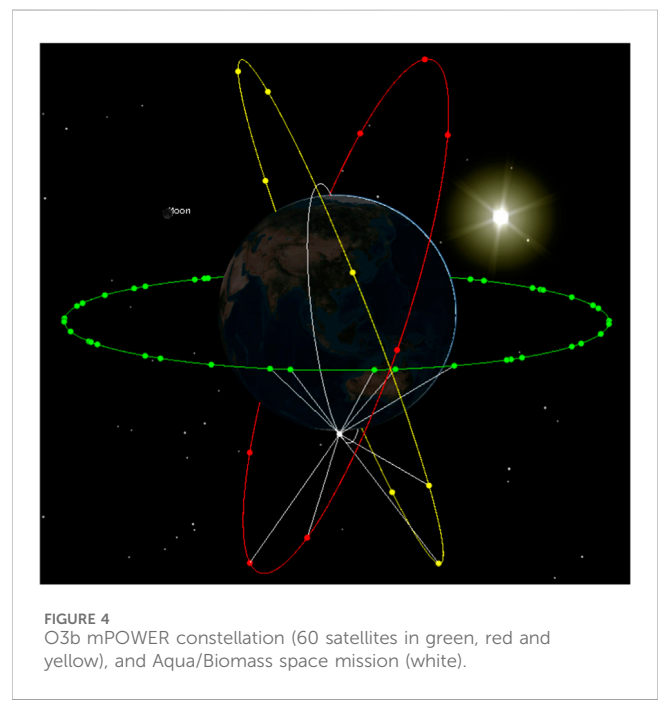

**FIGURE 4**
O3b mPOWER constellation (60 satellites in green, red and yellow, and Aqua/Biomass space mission (white).

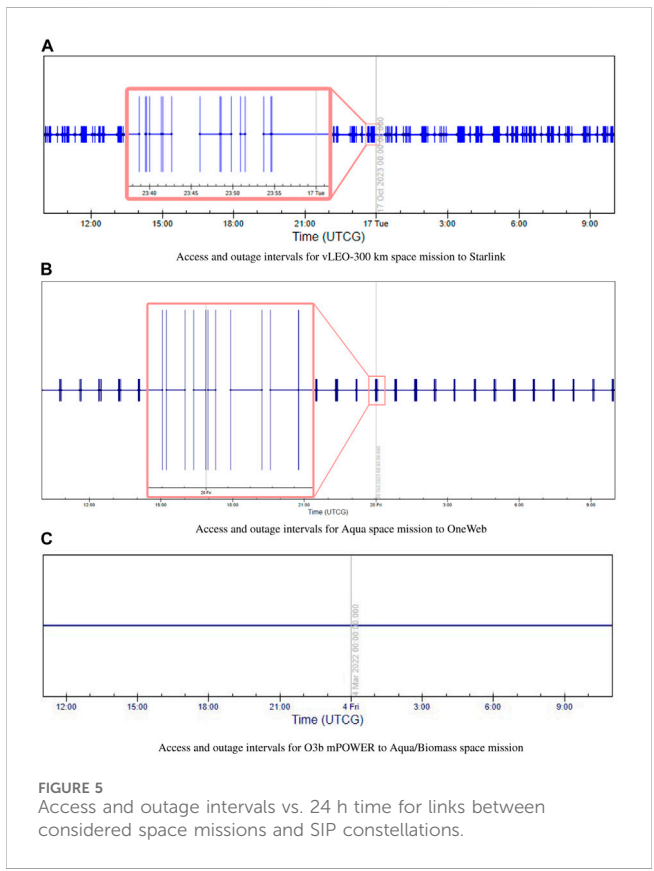

FIGURE 5
Access and outage intervals vs. 24 h time for links between considered space missions and SIP constellations.

Before concluding on the coverage results, it is worth noting here that another important metric, which is the percentage of satellites that provide a certain coverage duration, should be analyzed. This is because certain access periods are useless from a communication perspective since they are too short to establish a communication link or to perform the handover procedure when the link is already established. Of course, the required time either for the access phase or handover depends on the communication system and the adopted protocol. For example, it is evident that Starlink's minimum access duration of 4.62 s is not enough to establish communication but we wanted to emphasize that the other access durations without interruption, including the maximum ones, might be not fully exploited because it represents the access time from the constellation (i.e., a set of satellites) and not per satellite. For example, suppose a constellation has 100 side-by-side satellites with 2 s access per satellite (which is true for several satellites in Starlink and OneWeb constellations) which results in 200 s coverage time which appears to be good. However, the time per satellite is too short to be exploited for communication, and the *good* time coverage of the constellation is useless in this example. This is why it is important to assess the time coverage per satellite in the constellation to conclude the useful coverage time.

To assess the above metric, we provide in Figure 6 the empirical cumulative distribution function (eCDF) of the access duration per satellite, for each SIP constellation.

Now, assuming 20 s is a useful time to establish a communication. For Starlink and OneWeb, this means that space users would not be able to obtain connectivity at all times even though there is a coverage (see Figure 6). For example, 3.5% (i.e., 154 satellites) and 1.6% (i.e. 10 satellites) of the Starlink and OneWeb constellation satellites are useless for communication assuming the 20 s threshold. The discontinuity of communication with significant interruptions suggests that the best candidates for SIP are constellations orbiting at higher altitudes. O3b mPOWER with far more streched eCDF (at the bottom of the graph) all satellites in the constellation have enough time to guarantee communication.

## 3.3 Link budget and data rate performance

We provide in this sub-section an evaluation of the link budget and assessment of the performance for a hypothetical system composed of the selected Aqua space missions and the most suitable SIP system the O3b mPOWER. First, we provide the parameters of O3b mPOWER and space mission in Table 6. The space mission payload parameters are derived from Section 3.1. Parameters of two user terminals are considered here, the NanoSat and ThinPack Ka100 terminals. The former represents the case of a low-performance terminal while the

TABLE 5 Access intervals from space missions to SIP constellations.

| SIPs | Space mission | Min. Access duration w/o interruption [s] | Max. Access duration w/o interruption [s] | Total access duration [s] |
|---|---|---|---|---|
| **Starlink** | vLEO-300 km | 4.62 | 2099.32 | 73614.74 (85.20%) |
| **OneWeb** | Biomass | 22.40 | 5891.51 | 85847.54 (99.36%) |
| | Aqua | 6.69 | 2803.75 | 82991.99 (96.06%) |
| **O3b mPOWER** | Biomass | 86400 | 86400 | 86400 (100%) |
| | Aqua | 86400 | 86400 | 86400 (100%) |

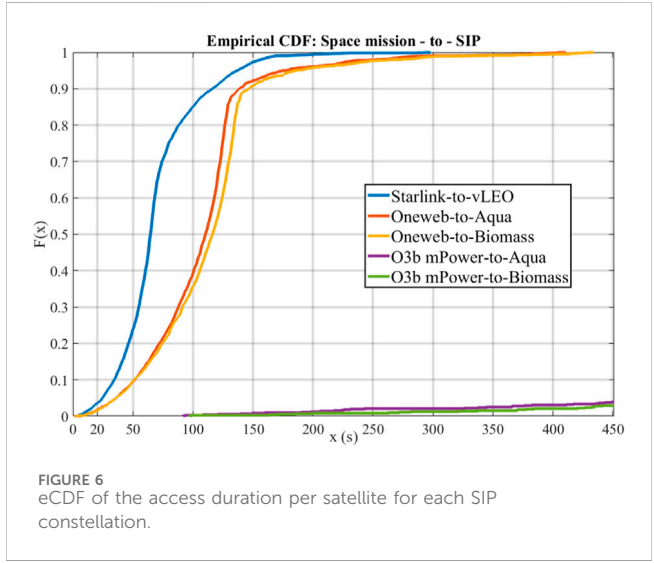

FIGURE 6
eCDF of the access duration per satellite for each SIP constellation.

TABLE 6 System parameters for numerical link budget evaluation.

| Parameters | O3b mPOWER | Space mission | |
|---|---|---|---|
| | | NanoSat | Ka100 |
| Satellite altitude | 8,062 km | 700 km | 700 km |
| Downlink transmission | | | |
| Frequency | Ka-band (i.e. 20 GHz) | | |
| Bandwidth | 100 MHz | | |
| EIRP | 49.7 dBW | NA | |
| G/T | NA | 2.2 dB/K | 13 dB/K |
| Uplink transmisson | | | |
| Frequency | Ka-band (i.e. 30 GHz) | | |
| Bandwidth | 4 MHz | | |
| EIRP | NA | 36.4 dBW | 46 dBW |
| G/T | 7 dB/K | NA | |

G/T, antenna Gain-to-noise-Temperature; EIRP, effective isotropic radiated power; NA, not available.

latter represents a high-performance one. We performed this choice to get lower and higher bound performance. The parameters of O3b mPOWER are derived from the FCC filling in (34). Figure 7 provides the link budget results for both uplink (UL, space mission to SIP link) and downlink (DL, SIP to space mission link) transmissions.

The obtained Energy symbol to Noise density ($E_s/N_0$) differs depending on the space mission payload characteristics (see Figures 7D, F). As it can be noted, the obtained $E_s/N_0$ is better in the DL compared to the UL. This is mainly due to the higher effective isotropic radiated power (EIRP) of the SIP's satellite compared to the space mission one. Also, the $E_s/N_0$ is time-variant due to the time-varying distance between the space mission satellite and SIP one. The maximum achievable $Es/N_0$ in the UL (resp. DL) is about 14.5 dB (resp. 18 dB) for Ka100-like payload, while for the NanoSat-like payload, it is about 5.5 dB (resp. 7.5 dB). On the other hand, the minimum $Es/N_0$ in the UL (resp. DL) is about 8 dB (resp. 11.5 dB) for Ka100-like payload and −1 dB (resp. 1 dB) for NanoSat-like payload. Since multi-users are competing for the UL resources, the $Es/N_0$ of the latter will be variable depending on the scheduled resources (i.e., bandwidth) from the provider.

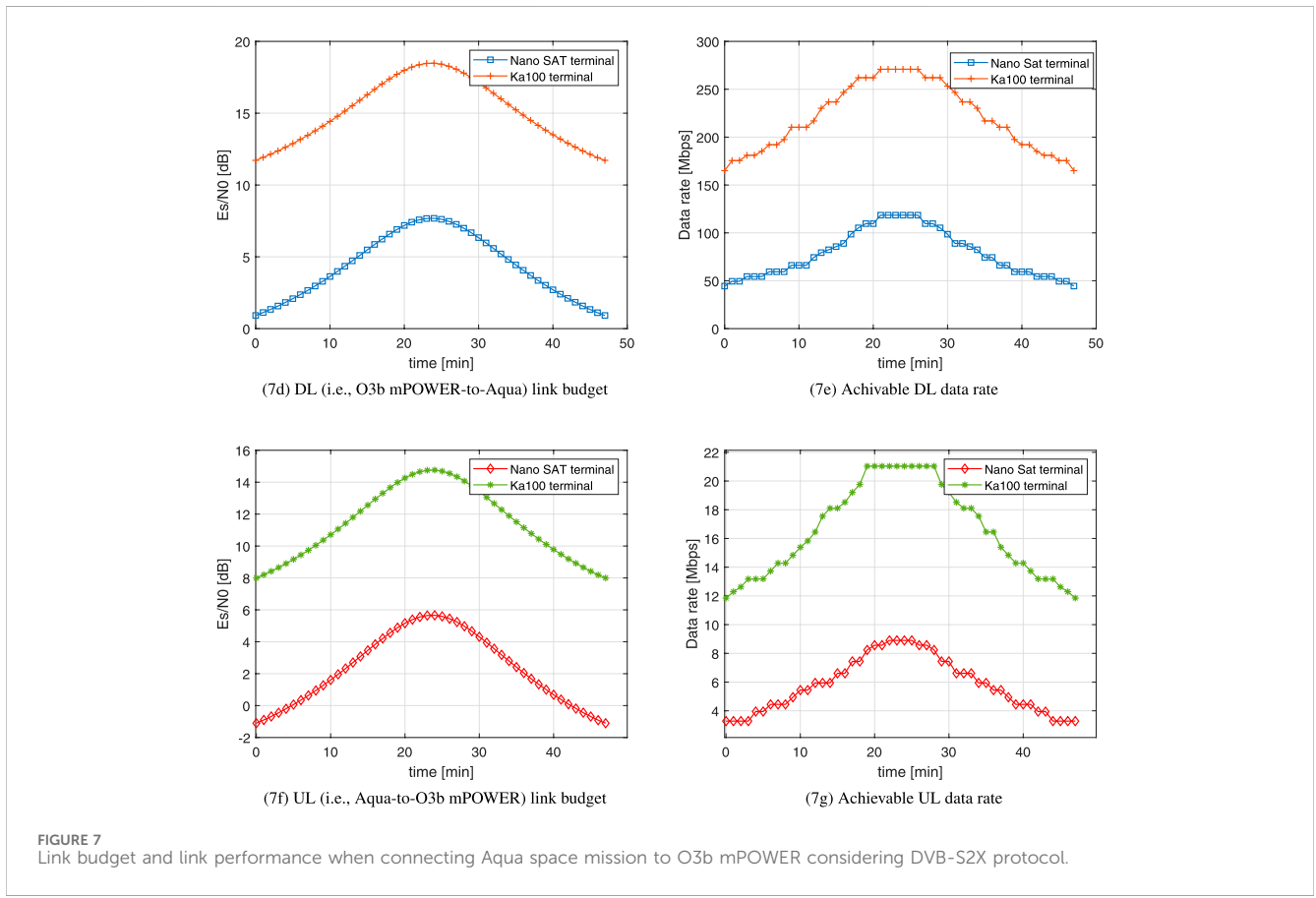

FIGURE 7
Link budget and link performance when connecting Aqua space mission to O3b mPOWER considering DVB-S2X protocol.

In order to quantify the achievable data rate, we assume here a digital video broadcasting second generation standard (DVB-s2x) protocol which is the most probable protocol used by O3b mPOWER system (Rao, 2022). Therefore, the data rate is obtained based on the DVB-s2x spectral efficiency (European Telecommunications Standards Institute, 2014) and the obtained $Es/N_0$s. More specifically, DVB-s2x protocol defines a set of modulation schemes and code rates (MODCODs) to be used to achieve a quasi error-free performance, i.e., packet error rate (PER) of $10^{-5}$. In such a configuration, the protocol changes the MODCOD when the $Es/N_0$ (i.e., SNR) changes to keep the same performance. Hence, all the data rate results presented below are obtained at a fixed PER of $10^{-5}$.

The results are depicted in Figure 7. For the DL, the achievable data rate is very high. It ranges from 165 Mbps to 270 Mbps for Ka100-like payload, and 50 Mbps to 118 Mbps for nanosat-like payload. For the UL, which is the most important link in our context[3], the achievable data rate ranges from 11 Mbps to 21 Mbps for Ka100-like payload and 3.2 Mbps to 8.8 Mbps for NanoSat-like payload. The obtained data rates are sufficient for many applications (e.g., video, data sharing) and can be compliant with space missions requirements. Note that the most important fact here is that these data rates can be permanently available and therefore real-time data exchange and tasking can be available for the connected space missions.

# 4 Legal and commercial challenges under the current spectrum usage policy

In order to understand the feasibility from a regulatory perspective of the envisaged space-based internet connectivity under the International Telecommunication Union ("ITU") regime, it is necessary to examine the current spectrum usage policy. In fact, to grant interference-free operations the spectrum is divided into bands and is allocated to specific services (Codding, 1994), subject to regular revision by the World Radiocommunication Conference ("WRC") (ITU, 2023). For the purpose of this study, we have based ourselves on the following assumptions, which are to be considered the most likely scenario to occur:

- the constellation has been notified at the ITU as a "service" by the competent Administration ("ADMIN 1") and the constellation provider has been licensed to provide services;
- the space mission has been notified at the ITU by a different Administration ("ADMIN 2") which has licensed the relevant space mission to be operated via the constellation;
- the constellation provider and the space mission operator operate their satellites respectively from the country of ADMIN 1 and country of ADMIN 2.

At this point, to establish communication between the SIP and the space mission in accordance with the ITU's regulatory framework, the system should either use (a) frequency filings for inter-satellite links assignments or (b) space-to-space allocations. In fact, the ITU's Radio Regulations ("RR"), the main legal instrument to regulate radiocommunication services and the utilization of the radio frequencies, divide satellite services in different categories[4] based on the different class of radiocommunication services that the satellites are planned to provide (ITU, 2021). For this purpose, the so-called "Table of Frequency Allocations" as per Article 5 of the RR lists the frequency bands allocated to the different satellite services, and in some cases also dictates directionality indicators of the transmissions (space-to-space, space-to-Earth and Earth-to-space) within the frequency bands allocated to a specific service, like, for example, the Space Research Satellite Service. From this it can be derived that beyond a proper inter-satellite link transmission covered under the framework of the Inter-Satellite Services, the radio frequency spectrum regime also envisages the possibility to establish a satellite-to-satellite link where there is an allocation in the space-to-space direction within the bandwidth allocated to a specific satellite service.

We observed that the constellation systems under investigation in this study (Starlink, OneWeb and O3b mPOWER) neither use the frequencies allocated at the moment to the Inter-Satellite Services nor can resort to space-to-space transmissions within the allocated frequencies for the satellite services they rely on[5], meaning Ka and Ku bands under the Fixed Satellite Services ("FSS"). It derives that at the state of the art, these systems could only file for an assignment under Article 4.4[6]. The latter is an assignment not in accordance with the Table of Frequency Allocations and Article 11.31 RR, a so-called "non-conforming assignment". These assignments do not enjoy the right to international recognition and must be operated on non-interference and non-protection bases. Moreover, assignments under Art. 4.4 impose additional obligations on the notifying administrations and national legislations may, to a different extent, discourage or prohibit non-conforming assignments (ITU, 2021). This means that satellite operators wishing to use non-conforming assignments are likely to encounter problems to be licensed at the national level (El-Moghazi et al., 2017). It becomes clear that recourse to Article 4.4 assignments is likely to have an impact on the actual exploitation of the service and might act as a deterrent for the envisioned operations.

---

3 Space mission mainly downloads data that were acquired during the mission.

4 See Article 1, Section III RR.

5 At the time of writing, the Table of Frequency Allocations does not allocate any space-to-space transmission to the FSS.

6 Article 4.4 RR: "Administrations of the Member States shall not assign to a station any frequency in derogation of either the Table Frequency Allocations in this Chapter or the other provisions of these Regulations, except on the express condition that such a station, when using such a frequency assignment, shall not cause harmful interference to, and shall not claim protection from harmful interference caused by, a station operating in accordance with the provisions of the Constitution, the Convention and these Regulations".

The situation just described might change according to the results of WRC-23 and the possible revision of the frequency allocations. More specifically, Agenda Item 1.17, set out by Resolution 773 (WRC-19), has attributed Working Party 4A the responsibility to determine the appropriate regulatory actions necessary for the provision of satellite-to-satellite links in specific frequency bands and to identify additional inter-satellite service allocations (ITU, 2019). Consequently, two possibilities might come to reality during WRC-23. We noted how the impact on the feasibility of space-based internet connectivity would be different if, on the one side, changes in WRC-23 would lead to (a) the allocation of Ka and Ku bands to the Inter-Satellite Services or, on the other hand, (b) the addition of space-to-space transmissions within the allocated frequency bands. In the case (a) above, the legal challenges stemming from an assignment under Article 4.4 RR would be either removed or at least mitigated, depending on whether Inter-Satellite Services are included as primary or secondary services in Ka and Ku bands[7] (I.-R. S. Groups, 2022). At the time of writing, the developments regarding Agenda Item 1.17 suggest that case (b) is the most probable outcome of WRC-23, with the introduction of space-to-space transmissions within the allocated frequency bands of the FSS. The main difference from the case (a) is that the satellite-to-satellite link must operate within a stricter regime in terms of requirements. Operational limitations derive from the "within the cone" concept of operations, meaning operations that must comply with the area of space already covered by the satellite filing under the FSS (I.-R. S. Groups, 2022). This means that if the constellation defines a specific cone as service area in its filing, the establishment of a satellite-to-satellite link between the constellation and the space mission under the FSS frequency band allocation cannot result in novel geometries that would extend the service beyond the declared service area. Consequently, we concluded that should these changes be implemented by WRC-23, the constellation, and the space mission shall comply with the following requirements:

- the constellation shall use the FSS frequency bands allocations;
- the satellite-to-satellite transmissions between the space mission and the constellation shall take place within the cone of coverage of the constellation;
- the space mission shall be at a lower orbital position than the constellation;
- the transmissions shall respect the directionality indicators as allocations.

Based on the analysis of the spectrum policy relevant to the proposed concept, we concluded that the current ITU regime applicable for the establishment of a satellite-to-satellite link between the constellation and the space mission (Article 4.4 RR assignments) is deemed not suitable for the provision of space-based internet connectivity. The lack of international recognition increases the risk of harmful interference with other operators with conforming assignments and therefore might impede or severely prejudice the actual use of these frequencies for satellite-to-satellite communication. For example, it might expose the space mission to sudden cuts or interruptions of the internet services provision, losing the advantage of a permanent/near permanent connection. In fact, the opportunistic access to the frequency spectrum is unreliable for commercial and public uses and therefore severely limits the feasibility of these operations (El-Moghazi et al., 2017). At the same time, the revision of the RR during WRC-23 as described above might constitute a turning point in the actual implementation of satellite communication via satellite internet service providers. In such a hypothesis, there would be an adequate legal protection and better premises for a continuous and efficient exploitation of satellite internet services. This would constitute a huge step forward in the provision of satellite internet services to space missions.

# 5 Conclusion

This paper evidences that the connection of space missions via satellie internet providers is beneficial and most importantly is practically possible in the near future. It was found that the most probable candidate that can take advantage of the proposed concept is Earth observation missions because of (1) their overwhelming need to transfer large amounts of data, (Vrancken et al., 2014), their preference to transfer data quickly to enable time critical observations and missions, (Al-Hraishawi et al., 2023), their orbital compatibility with mega-constellations (i.e., their low orbits) and (Idrsspace, 2023) their abundant mass of users. Our investigation showed that commercially available terminals have a great potential to be used on LEO space mission spacecraft. Their overall maturity allows their integration into satellites after the delta design work on the thermal management and other detailed review-of-design verification has been carried out. It turned out that O3b mPOWER is the most suitable constellation for the proposed concept, followed by OneWeb, thanks to their relatively high altitude. Starlink may also be suitable for future vLEO missions. The evaluation showed that for a hypothetical system composed of Aqua space mission and O3b mPOWER, can reach up to 21 Mbps data rate in the UL considered as the most critical link. As per the legal aspects, the study focused on the current legal regime at the ITU for the establishment of the satellite-to-satellite link, and the necessary conditions for compliance with the regulatory framework. It was found that at the-state-of-the-art, the systems under investigation could only rely on a flexible use of the spectrum allowed on a non-interference and non-protection basis, the so-called Article 4.4 assignments. The lack of international protection jeopardizes the operations at issue. This commercial and operational disadvantage could be bridged by the possible changes to the Table of Frequency Allocations during the 2023 World Radiocommunication Conference.

---

7  Frequency bands can be allocated to shared bands. Services in these shared bands can be divided into primary services or secondary services. Primary services enjoy superior rights than secondary ones. As a consequence, the latter shall neither cause harmful interference to nor claim protection from the former.

## Data availability statement

The original contributions presented in the study are included in the article/Supplementary Material, further inquiries can be directed to the corresponding author.

## Author contributions

HC: Conceptualization, Methodology, Project administration, Validation, Writing–original draft, Writing–review and editing, Formal Analysis, Supervision. OK: Data curation, Investigation, Writing–original draft. AG: Formal Analysis, Investigation, Methodology, Resources, Software, Visualization, Writing–original draft, Writing–review and editing, Validation. JT: Formal Analysis, Investigation, Methodology, Resources, Writing–original draft, Writing–review and editing, Data curation. CV: Data curation, Investigation, Writing–original draft, Writing–review and editing. FZ: Formal Analysis, Investigation, Writing–original draft. PP: Formal Analysis, Investigation, Writing–original draft. MH: Supervision, Writing–review and editing. SC: Project administration, Funding acquisition, Validation, Writing–review and editing.

## Funding

The author(s) declare financial support was received for the research, authorship, and/or publication of this article. This work was supported by European Space Agency in the frameworks of the European Space Agency Project "Satellite Communications via Satellite Internet Service Providers".

## Conflict of interest

OK and CV was employed by SES S.A.

The remaining authors declare that the research was conducted in the absence of any commercial or financial relationships that could be construed as a potential conflict of interest.

## Publisher's note

## Supplementary material

The Supplementary Material for this article can be found online at: https://youtu.be/Xj5eRmco770